\documentclass[a4paper,11pt]{article}
\pdfoutput=1 % if your are submitting a pdflatex (i.e. if you have images in pdf, png or jpg format)
\usepackage{jcappub}

\usepackage{amsmath,amssymb,amsbsy,amstext, amsthm, simplewick, amsfonts}

\usepackage{graphicx}

\usepackage{siunitx}
\usepackage{upgreek}
\usepackage{framed}
\usepackage{wrapfig}
\usepackage{multirow}
\usepackage{bbm}

\usepackage[absolute,overlay]{textpos}

\begin{document}

\begin{textblock*}{5cm}(18.46cm,1cm)  % 位置可微调
    % \small YITP-25-67
    
\end{textblock*}

\arxivnumber{2506.06797} 

% \preprint{APS/123-QED}

%\title{Reheating, Radiation and Dark Matter from Dual Primordial Black Holes}% Force line breaks with \\

\title{Primordial black holes save $R^2$ Inflation}

\author{Xinpeng Wang$^{1,2,3}$}
\emailAdd{xinpeng.wang{}@ipmu.jp}

\author{Kazunori Kohri$^{1,4,5,6}$}
\emailAdd{kazunori.kohri{}@gmail.com}

\author{Tsutomu T. Yanagida$^{1,7}$}
\emailAdd{tsutomu.tyanagida{}@gmail.com}
% \author{Ying-li Zhang$^{5,6,7,8}$}
% \email{yingli{}@tongji.edu.cn}

\affiliation{
 $^1$ Kavli Institute for the Physics and Mathematics of the Universe (WPI), The University of Tokyo Institutes for Advanced Study, The University of Tokyo, Chiba 277-8583, Japan\\
 $^2$ Department of Physics, Graduate School of Science, The University of Tokyo, Tokyo 113-0033, Japan\\
 $^3$ School of Physics Science and Engineering, Tongji University, Shanghai 200092, China\\
 $^4$ Division of Science, National Astronomical Observatory of Japan,2-21-1 Osawa, Mitaka, Tokyo 181-8588, Japan\\
 $^5$ School of Physical Sciences, Graduate University for Advanced Studies (SOKENDAI), 2-21-1 Osawa, Mitaka, Tokyo 181-8588, Japan\\
 $^6$ Theory Center, IPNS, KEK, 1-1 Oho, Tsukuba, Ibaraki 305-0801, Japan\\
 $^7$ Tsung-Dao Lee Institute \& School of Physics and Astronomy, Shanghai Jiao Tong University, Pudong New Area, Shanghai 201210, China}
 % \\
 % $^7$Institute for Advanced Study of Tongji University, Shanghai 200092, China\\
 % $^8$Center for Gravitation and Cosmology, Yangzhou University, Yangzhou 225009, China}
% \affiliation{%
%  Authors' institution and/or address\\
%  This line break forced with \textbackslash\textbackslash
% }%

% \collaboration{MUSO Collaboration}%\noaffiliation

% \author{Charlie Author}
%  \homepage{http://www.Second.institution.edu/~Charlie.Author}
% \affiliation{
%  Second institution and/or address\\
%  This line break forced% with \\
% }%
% \affiliation{
%  Third institution, the second for Charlie Author
% }%
% \author{Delta Author}
% \affiliation{%
%  Authors' institution and/or address\\
%  This line break forced with \textbackslash\textbackslash
% }%

% \collaboration{CLEO Collaboration}%\noaffiliation

\date{\today}% It is always \today, today,
             %  but any date may be explicitly specified

\abstract{
In light of the latest Planck and Atacama Cosmology Telescope (P-ACT) joint results on the primordial scalar power spectrum, we show that the $R^2$ inflation model extended with a non-minimally coupled scalar field $\chi$—namely the $\chi$-extended $R^2$ inflation model—can naturally accommodate a larger spectral index $n_s$ and a small positive running $\alpha_s$ at cosmic microwave background (CMB) scales, both of which are consistent with the latest P-ACT constraints. This is because the $\chi$ field contributes a blue-tilted component to the primordial power spectrum, which both modifies the large-scale power and, as a result, significantly enhances power on small scales. The deviation of the $n_s$ and $\alpha_s$ from the single field $R^2$ inflation is related to the non-minimal coupling constant $\xi$. The consequent enhancement in the primordial power spectrum can be large enough to lead to the formation of primordial black holes (PBHs) of mass $\lesssim 10^{20}\mathrm{g}$ as dark matter candidates. Furthermore, future observations of the small-scale power spectrum, CMB spectral distortions, and stochastic gravitational waves will provide decisive tests of this model and its predictions for PBHs. We stress its strong connection to the seesaw mechanism for the generation of the observed small masses.}

%\keywords{Suggested keywords}%Use showkeys class option if keyword
                              %display desired
\maketitle

%\tableofcontents

\section{Introduction}
\noindent

The $R^2$ inflation~\cite{Starobinsky:1980te,Mukhanov:1981xt}, also known as the Starobinsky inflation, has long been regarded as one of the most successful inflationary models, both theoretically well-motivated and perfectly fits the observations.
Its predictions for the spectral index $n_s$ of the primordial power spectrum and the tensor-to-scalar ratio $r$ are in excellent agreement with the Planck 2018~\cite{Planck:2018jri} and BICEP/Keck observations. The model also gives a precise prediction of a negative running of spectral index $\alpha_s$, which is also consistent with the Planck results \footnote{According to the Planck 2018 result, $n_s=0.9649\pm0.0042$ (68\% CL),$\alpha_s\equiv dn_s/d\ln k= -0.0041 \pm 0.0067$ (68\% CL). The tensor-to-scalar ratio is constrained to $r<0.032$ (95\% CL) by joint Planck and BICEP/Keck observations~\cite{Tristram:2021tvh}.}.

However, a newly released result on the primordial power spectrum from the Atacama Cosmology Telescope (ACT) DR6~\cite{AtacamaCosmologyTelescope:2025blo, AtacamaCosmologyTelescope:2025nti} reported a slightly larger spectral index. Combined with Planck data at large-to-intermediate scales (denoted as P-ACT) and together with CMB lensing and BAO data from DESI~\cite{DESI:2024uvr} (denoted as P-ACT-LB), the analysis yields a new constraint on the spectral index: $n_s=0.9743 \pm 0.0034$ (68\% CL) at the CMB pivot scale $k_{\mathrm{CMB}}=0.05~\mathrm{Mpc}^{-1}$~\cite{AtacamaCosmologyTelescope:2025blo}. This result appears to disfavor the $R^2$ inflation as illustrated in figure \ref{nsrcase2}. In addition, the error bar for the running of the spectral index $\alpha_s$ is tightened in the P-ACT-LB result, which suggests a slight positive running of the spectral index
$\alpha_s=0.0062\pm 0.0052$ (68\% CL)~\cite{AtacamaCosmologyTelescope:2025nti}. The constraint on $\alpha_s$ is not consistent with most of the single-field models which involve a convex potential, including the $R^2$ model. 
Several works~\cite{Asaka:2015vza,Koshelev:2022olc,Cheong:2019vzl,Kim:2025dyi, Gialamas:2025ofz,Haque:2025uis,Yogesh:2025wak,Addazi:2025qra, Drees:2025ngb,Frolovsky:2025iao} discussed extensions to the single-field $R^2$ model that may resolve this inconsistency.

In this paper, we present a simple extension to the $R^2$ model by introducing a non-minimally coupled scalar field $\chi$ (this form is also adopted in $R^2$-Higgs model~\cite{Salvio:2015kka, Salvio:2016vxi, He:2018gyf, Cheong:2019vzl, Cheong:2022gfc, Wang:2024vfv,Kim:2025dyi}) that may provide a better fit to the newly released data, thus potentially "save" the $R^2$ model via a hybrid-like inflationary scenario. The $\chi$ field, acting as a waterfall field, introduces a blue-tilted contribution to the total primordial power spectrum, thereby leading to a positive shift in both the spectral index $n_s$ and its running $\alpha_s$. Notably, this blue-tilted contribution peaked at small scales can trigger the gravitational collapse of overdense regions into primordial black holes (PBHs)~\cite{Zeldovich:1967lct, Hawking:1971ei, Carr:1974nx, Meszaros:1974tb, Khlopov:1985fch} during radiation domination, which could constitute all cold dark matter. 

Therefore, we also discuss the possible hint from the P-ACT results to the small-scale features in the primordial power spectrum. In particular, we find that the scenarios involving the formation of primordial black holes with mass below $10^{20}\mathrm{g}$ through the enhanced small-scale power spectrum, accounting for all of the dark matter, show good agreement with the above latest data. The growth of the power spectrum in our model also gives rise to distinctive, potentially detectable signatures in both the $\mu$-distortion and the stochastic gravitational wave background, which can serve as distinguishing features in future surveys like PIXIE~\cite{Kogut:2011xw,Abitbol:2017vwa,Tagliazucchi:2023dai}, LISA~\cite{LISA:2017pwj,LISACosmologyWorkingGroup:2024hsc,LISACosmologyWorkingGroup:2025vdz}, etc.

A crucial point in our model is the presence of a new boson field $\chi$ which has an $Z_2$ invariant potential where the $\chi$ is odd under the discrete symmetry. We identify it with the $Z_4$ symmetry whose breaking generates the Majorana masses $M_i$ for the right-handed neutrinos $N_{Ri}~(i=1-3)$ ~\cite{Kawasaki:2023mjm}. Here, the $\chi$ and $N_{Ri}$ have the $Z_4$ charges $+2$ and $-1$, respectively. This $Z_4$ is free from the Dai-Freed anomaly~\cite{Dai:1994kq} and hence it is a perfect symmetry~\cite{Witten:2015aba,Yonekura:2016wuc} in the standard model. The breaking scale of the $Z_4$ is $\langle\chi\rangle\simeq 10^{12}$ GeV which gives the Majorana masses $M_i\sim g_i\langle\chi\rangle$.  Those masses are very much consistent with the observed small masses for the neutrinos~\cite{Minkowski:1977sc,Yanagida:1979as,Yanagida:1979gs,Gell-Mann:1979vob}. 

The remainder of this paper is organized as follows. In section 2, we describe the setup and analyze the field dynamics. Section 3 presents the semi-analytical treatment of the enhancement in the curvature perturbations, and its implications for CMB observables such as the spectral index $n_s$
, its running $\alpha_s$
, and the tensor-to-scalar ratio $r$, as well as for the formation of primordial black holes (PBHs). section 4 compares our numerical results with the latest CMB measurements, highlighting the parameter regions consistent with current constraints. Finally, section 5 provides a summary and discussion of our main conclusions.

\section{The model}
We consider the action in which $R^2$ gravity is non-minimally coupled to a scalar field $\chi$~\cite{Pi:2017gih, He:2018gyf, Cheong:2019vzl, Cheong:2022gfc, Wang:2024vfv},
% with a slightly broken $Z_2$ symmetry characterized by $\chi_0$~\footnote{The $Z_2$ breaking term is introduced in hybrid models to resolve the overproduction of domain walls and avoid the unacceptably huge quantum fluctuations~\cite{Braglia:2022phb, Wang:2024vfv}},
\begin{equation}\label{Jordan}
\begin{aligned}
    S_{J}=\int d^4x\sqrt{-g}\ \bigg[\frac{M_{\mathrm{pl}}^2}{2}f(R,\chi)-\frac{1}{2}g^{\mu\nu}\partial_\mu\chi\partial_\nu\chi-V(\chi)\bigg],
\end{aligned}
\end{equation}
where $M_{\mathrm{pl}}\equiv(8\pi G)^{-1/2}$, $f(R,\chi)$ is given by \footnote{See an extension to $R^3$ in \cite{Kim:2025dyi}}
\begin{align*}\label{fdef}
    &f(R,\chi)=R+\frac{R^2}{6M^2}-\frac{\xi R}{M_{\mathrm{pl}}^2}(|\chi|-\chi_0)^2,
\end{align*}
and $V(\chi)$ is a quartic potential,
\begin{align*}
    &V(\chi)= \frac{1}{4}\lambda\left(\chi^2-v^2\right)^2,
\end{align*}
where $v=\pm m/\sqrt{\lambda}$ is the vacuum expectation value of $\chi$, and $m$ is the mass of $\chi$.
Notice that the above action is invariant under the $Z_4$ symmetry under which the $\chi$ transforms to $-\chi$. The presence of $|\chi|$ is not a serious problem, since we consider it to be the low-energy action in an effective field theory (EFT). In fact, the $|\chi|$ dependent terms often appear at the EFT level when such a boson field couples to non-abelian gauge theories. (See \cite{Harigaya:2012pg} for example.) 

It is well-known that the $R^2$ term introduces a scalar degree of freedom to the theory. Introducing an auxiliary field $\psi$, the action \eqref{Jordan} is equivalent to,
\begin{equation}
    \begin{aligned}
     S_{\psi}=\int d^4x\sqrt{-g}\ \bigg[\frac{M_{\mathrm{pl}}^2}{2}RF(\psi,\chi)-\frac{1}{2}g^{\mu\nu}\partial_\mu\chi\partial_\nu\chi-V(\chi)-U(\psi)\bigg],
\end{aligned}
\end{equation}
where 
\begin{align}
   F(\psi,\chi)&=1+\frac{\psi}{6M^2}-\frac{\xi}{M_{\mathrm{pl}}^2}(|\chi|-\chi_0)^2\\
    U(\psi)&=\frac{M_{\mathrm{pl}}^2\psi^2}{12M^2}
\end{align}
To bring the above action into the Einstein frame, we perform the conformal transformation  $\tilde g_{\mu\nu}=\Omega^2 g_{\mu\nu}$~\cite{DeFelice:2010aj} for the choice of $\Omega^2=F$, where the $\Omega^2$ is the conformal factor, and a tilde represents quantities in the Einstein frame. Under transformation, the action~\eqref{Jordan} can be written in the Einstein frame,
% can be transformed into the Einstein frame~\cite{DeFelice:2010aj} as 
\begin{equation}
    \begin{aligned}\label{EinsteinS}
    S_{\mathrm{E}}=\int d^4 x\sqrt{-\tilde g}\left[\frac{M_{\mathrm{pl}}^2}{2}\tilde R-\frac{1}{2}\tilde g^{\mu\nu}\partial_{\mu}\phi\partial_{\nu}\phi-\frac{1}{2}F^{-1}\tilde g^{\mu\nu}\partial_\mu\chi\partial_\nu\chi-U(\phi,\chi)\right],
\end{aligned}
\end{equation}
where the scalar field $\phi$ is defined by
% the $R^2$ term converted into a scalar degree of freedom $\phi$, given by
\begin{align*}
    \phi &\equiv M_{\mathrm{pl}} \sqrt{\frac{3}{2}} \ln F,
\end{align*}
equivalently, we define $F(\phi)\equiv \exp(\sqrt{2/3}\phi/M_{\mathrm{pl}})$
and with an effective potential written as
\begin{align*}
  &  U(\phi,\chi)
\equiv\frac{3}{4}M^2M_{\mathrm{pl}}^2W^2(\phi,\chi)+\frac{V(\chi)}{F(\phi)^2},\\
& W(\phi,\chi)\equiv 1-\frac{1}{F(\phi)}\left[1-\xi\left(\frac{|\chi|-\chi_0}{M_{\mathrm{pl}}}\right)^2\right].
\end{align*}
From now on, we work in the Einstein frame and set $\hbar=c=M_{\mathrm{pl}}=1$, adopting natural units for notational simplicity.

% T.T.Y. [Can people understand the above? Eq.(1) contains R and chi only, but Eq.(2) contains $\tilde R$, phi and chi. It looks that you increased the number of the fields. If not we need an explanation why not.]
% \xw{Please find my explanations above. In general people work in f(R) theory could understand, based on the idea that the f(R) is equivalent to scalar-tensor theory with 1 scalar degree of freedom.}

% T.T.Y. [Why $|\chi|$ has disappeared in the potentials U and W ?]
% \xw{I overlooked the action here, I added the $|\chi|$ back.}

In this framework, a hybrid-like inflation can be realized in the composition of two separated stages of effective single-field inflation — one dominated by $\phi$, the other by $\chi$ — separated by an intermediate transition stage, as described below:

\begin{figure}[!htbp]
\centering
% \includegraphics[width=.48
% \textwidth]{transferfuncnew1.pdf}
\includegraphics[width=.6\textwidth]{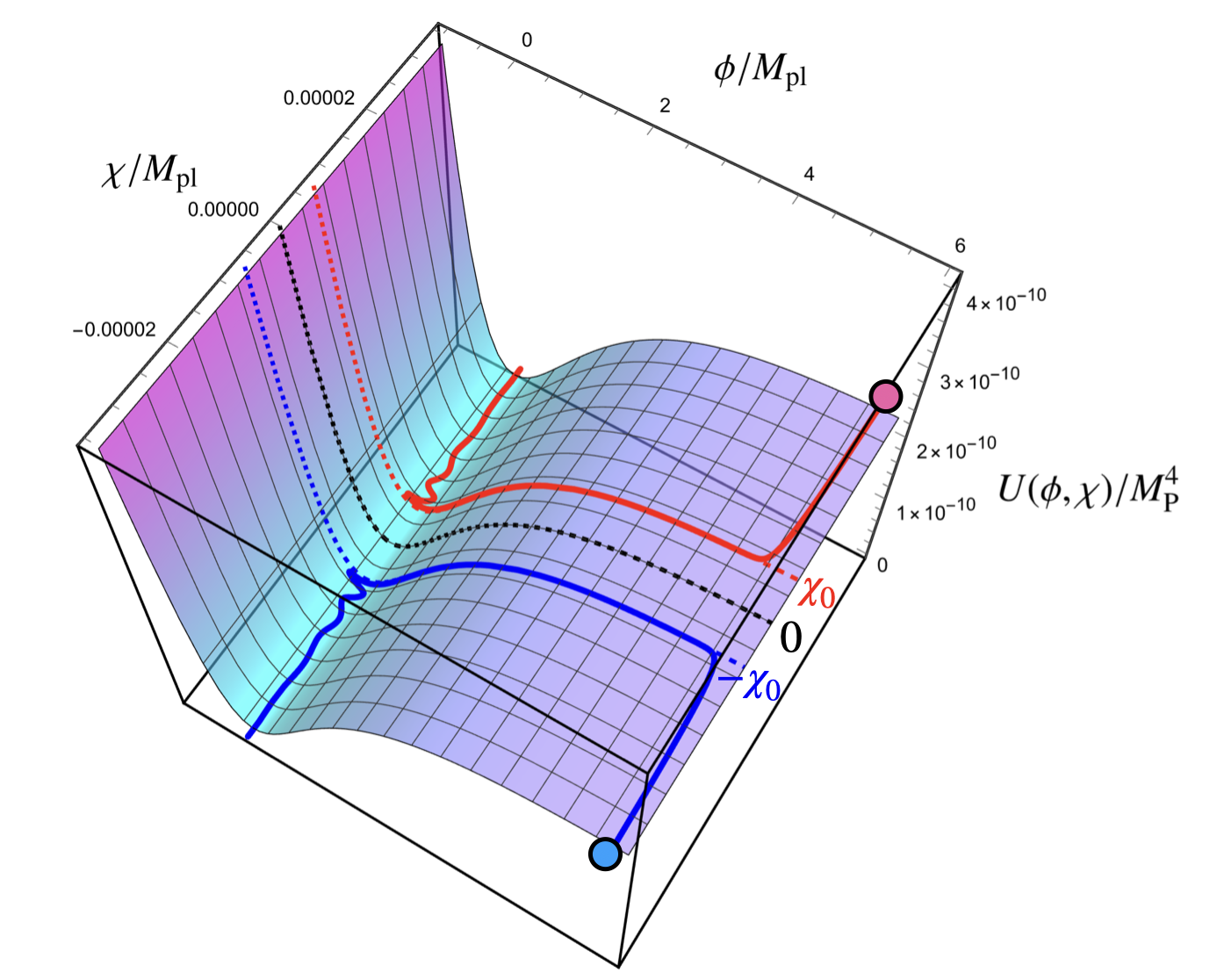}
\caption{\label{potentialshape}The shape of potential and the background solution with different initial conditions. The solid red and blue lines show the background trajectory obtained by numerically solving the background equations of motion with different initial conditions $\chi_{\mathrm{ini}}>0$ (red) and $\chi_{\mathrm{ini}}<-0$ (blue).}
\end{figure}

\begin{itemize}
    \item St-1: \textbf{Effective single field $R^2$ inflation.} 
    Initially, the inflaton slow-rolls from the plateau of the potential along the $\phi$ direction. In the large $\phi$ limit (equivalent to the large $F(\phi)$ limit), the potential becomes
\begin{equation}
    \begin{aligned}
    U(\phi,\chi)=&\frac{3M^2}{4}\\+&F(\phi)^{-1}\frac{3M^2}{2}\left(-1+\xi(|\chi|-\chi_0)^2\right)\\
    +&\mathcal{O}(F(\phi)^{-2})V(\chi)
\end{aligned}
\end{equation}
Given the large initial value of $\phi$, the spectator field $\chi$ acquires an effective mass approximately given by $m_{\chi} ^2\approx 3M^2\xi(1+\mathcal{O}(F(\phi)^{-1}))$. 
We assume that the initial value of $\chi$ is close to its potential minimum $\pm\chi_0$ at the beginning of inflation. During this stage, when $m_{\chi} ^2> 9H^2/4\approx 9M^2/16$, or equivalently $\xi>3/16$, $\chi$ is considered a heavy field and behaves as an underdamped oscillator. In contrast, when $m_{\chi} ^2< 9H^2/4$, $\chi$ is light and behaves as an overdamped oscillator; note that in this case, $\phi$ remains the lightest field. In each of the cases, the scalar field $\chi$ is settled down to one of its potential minimum ($\chi_0$ or $-\chi_0$, $\chi_0$ is taken for the computations in the following section) before this stage of inflation reaches an end.

% T.T.Y. [\chi_0$ is NOT vev. It is the potential minimum during the first sage of the inflation. Is it correct?]
% \xw{You are correct. I changed the text.}

% $\\$

%     T.T.Y. [Where is $R^2$ term ?]
% \xw{The $R^2$ term is converted to $\phi$ as we explained before, as long as $\xi(|\chi|-\chi_0)^2$ term is negligible, compared to $F(\phi)$, the inflation looks like a $R^2$ inflation. $R^2$ term is not appearing since we are working in the Einstein frame, where we write the theory as a scalar-tensor theory in Einstein gravity. }$\\$
%     T.T.Y. [Is it clear the plateau exists in a large phi/MPL region? And F is dimensionless. Why can we neglect 1/F effect in $m^2_\chi$ ?]
% \xw{Let me write down the potential for big $\phi/M_{pl}$ in order to make the statement clear. Since $F(\phi)=\exp({\sqrt{3/2}\phi/M_{pl}})$ is huge during this stage, so we can neglect.}$\\$
% T.T.Y.[I meant that you did not write the coefficient of 1/F in $m_\chi^2$ and hence I am not sure we can neglect 1/F term.]
%     T.T.Y. [We should mention where is the chi during the phi inflation.]
% \xw{Let me comment.}

    \item \textbf{Transition.} 
    The first stage of inflation ends when the inflaton $\phi$ reaches the local potential minimum in the $\phi$ direction. Simply put, the universe undergoes a short but complete version of effective single field $R^2$ inflation before this moment of transition.
    During this phase, $F$ approaches unity when $\phi$ reaches its VEV. The potential at this stage is given in the limit of $|\chi|\rightarrow \chi_0$, $\phi\rightarrow0$ and $|\chi_0|\ll m/\sqrt{\lambda}$, the potential is therefore approximately given by:
    \begin{align}
        U(\phi,\chi)\approx (\lambda/4)(\chi_0^2-v^2)^2\approx \frac{m^4}{4\lambda}-\frac{1}{2}m^2\chi^2
    \end{align}
    As shown above, the boson field $\chi$ becomes unstable $(m_{\chi}^2\approx-m^2< 0)$. Therefore, the inflaton dynamics begin to turn toward the $\pm\chi$ direction. 
    It is worth mentioning that $\chi$ field starting from a initial condition away from $0$ steps down the potential during the first stage of the $R^2$ inflation and hence the $Z_4$ symmetry is already broken for the background trajectory, or more precisely,inflation selects a vacuum branch during the early stage (see in figure \ref{potentialshape}).
    When fixing $|\chi_0|^2\gg\langle\delta\chi^2\rangle$ ~\cite{Sakai:1995nh,Braglia:2022phb}, the overproduction of domain walls (DWs) can be avoided, which usually happens in hybrid inflation models with a discrete $Z_2$ symmetry. 

    % % T.T.Y. [Normally, the inflation ends before the inflaton reach the minimal of its potential.]
    % % \xw{Here the potential minimal is only a local minimal for $\phi$, not the overall minimal.}

    % % T.T.Y. [The statement is un-understandable.]

    % T.T.Y. [It is very important to mention that the $\chi$ steps down neat by $\chi_0$] or $-\chi_0$ during the first stage of the R^2 inflation and hence the $Z_4$ is already broken. This is the reason why we do not have domain walls after the end of the total inflation.]
    
    \item St-2: \textbf{Effective single field $\chi$ inflation.} The second stage of inflation starts along $\chi$ direction from $\pm\chi_0$ (the sign will depend on the initial condition for $\chi$), with the heavy field $\phi$ stabilized in its potential minimum. We may observe that $\phi$ is still evolving during this stage, but it is simply due to the varying $\chi$ causing the movement of the potential minimum. The time evolution of $\phi$ is not violating the effective single field slow condition, since the isocurvature is heavy and decays out very rapidly. The second stage of inflation ends as an effective single field inflation of $\chi$. To have a sizable number of efolds during St-2, $v\gg \chi_0$ is required. 
At the same time, eternal inflation may happen during waterfall region due to the flatness of the potential along the background trajectory, which gives another constraint on $\chi_0$. To avoid eternal inflation, the classical displacement of the field within one Hubble time should be much larger than the quantum fluctuation, requiring
    \begin{align}
\Delta\chi_{\mathrm{cl}}\equiv\frac{\left|\dot\chi\right|}{H}\gg\sqrt{\langle\delta\chi^2\rangle},
    \end{align}
    where the typical size of inflaton quantum $\sqrt{\langle\delta\chi^2\rangle}\sim H/(2\pi)$ in the superhorizon limit.
    
    Combining with conditions to avoid DW overproduction and eternal inflation, the constraints for the parameters is given
    % \begin{align}
    % m^2/\lambda\gg\chi_0^2\gg M^2/(64\pi^2).
    % \end{align}
    \begin{align}
        \frac{m}{\sqrt{\lambda}}\gg|\chi_0|\gg  \max\left\{\frac{M}{4\pi},\frac{m^4}{16\sqrt{3}\pi\lambda^{3/2}}\right\}
    \end{align}

\end{itemize}

As we explained above, the scalar field $\chi$, acting as a spectator field (or equivalently, the isocurvature component) during the $R^2$ inflation, dominates the inflation during St-2. Such background evolution, commonly observed in hybrid inflation models~\cite{Gong:2010zf, Lyth:2012yp,Clesse:2015wea} with an initially heavy isocurvature field $\chi$, can induce an enhancement in the comoving curvature power spectrum steepest to $k^3$ and thus lead to PBH and GW production.

% T.T.Y. [I don't like the name of scalaron. What is wrong to write scalar boson $\chi$ ?]\xw{Indeed the scalar boson is correct.}
% T.T.Y. [What is the effective R^2 inflation above? Do you mean that "the first stage of the R^2 inflation?]
% \xw{Correct. Since the first stage of inflation is effective single field inflation (although we are considering two fields, only 1 field is dominating.), so I writed down effective single field $R^2$ inflation.}

A similar scenario arises in the curvaton model~\cite{Moroi:2001ct,Enqvist:2001zp, Lyth:2002my, Moroi:2005kz, Ichikawa:2008ne}. In contrast to the waterfall field in hybrid inflation, the curvaton remains effectively massless and nearly frozen during inflation. It dominates the universe once the Hubble parameter becomes smaller than the curvaton mass after the end of inflation. In such cases, the comoving curvature power spectrum contributed by curvaton field perturbations that decays later should be close to scale-invariant, depending on the curvaton's effective mass during inflation.

Notably, both mechanisms are encompassed within the rich phenomenology of the $\chi$-extended $R^2$ model. 
During the first stage of inflation, the effective mass of $\chi$ is given by $m_{\chi} ^2=F\partial^2 U/\partial\chi^2\sim 3M^2\xi$.  Since the Hubble parameter in this early stage is approximately $\sim M^2/4$, the ratio of $m_\chi^2$ to $H^2$ turns out to solely depend on the coupling constant $\xi$. 

Thus, by varying the coupling constant $\xi$ from 0 to values $\geq 3/16$, the model allows the contribution from $\chi$ to the adiabatic power spectrum to evolve from nearly scale-invariant ($k^0$) to steeply blue-tilted ($k^3$), as explained detailedly in the following section (also shown in~\cite{Wang:2024vfv}).

% Therefore, in principle, the model allows the power spectrum to grow from nearly scale-invariant ($k^0$) to steeply blue-tilted ($k^3$) by varying the coupling constant $\xi$ from $0$ to $\gtrsim 3/16$~\cite{Wang:2024vfv}. 

Consequently, if a significant enhancement in the power spectrum occurs at small scales, a steeper growth helps to localize the feature, keeping it well separated from large-scale modes. On the other hand, when the spectrum grows more gently, the enhancement at small scales may reshape the large-scale spectrum.
In the following, we present the $\xi$-dependence of the primordial power spectrum using $\delta N$ formalism.

\section{Semi-analytical power spectrum}
The $\delta N$ formalism relates the comoving curvature perturbation $\mathcal{R}$ to the field perturbations, which are evaluated on spatially flat slicing at superhorizon scales~\cite{Kodama:1984ziu,Starobinsky:1985ibc,Salopek:1990jq,Sasaki:1995aw}. Here, $N$ denotes the number of e-folds from the flat slicing to the comoving slicing at the end of inflation, and $\delta N$ represents the perturbation in $N$ induced by the superhorizon field fluctuations.

\begin{figure}[!htbp]
\centering
% \includegraphics[width=.48
% \textwidth]{transferfuncnew1.pdf}
\includegraphics[width=.6\textwidth]{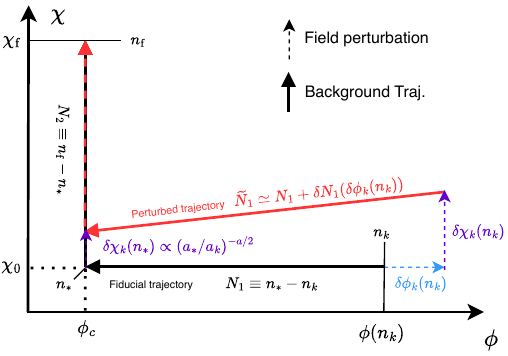}
\caption{\label{simplifiedtraj}A schematic diagram of the simplified trajectory. We assume that the transition stage occurs instantaneously at $n_*$, therefore the background evolution consists of two separate phases of effective single-field inflation along $\phi$ and $\chi$ directions, respectively. The fiducial background trajectory is indicated by black solid arrows, while the perturbed trajectory is shown with red solid arrows. 
For the perturbed trajectory, the boundary condition is perturbed by the field perturbations $\delta\phi_k$ (dashed blue arrow) and $\delta\chi_k$ (dashed purple arrows) in the spatially flat gauge at the horizon exiting stage of mode $k$, denoted by $n_k$.}
\end{figure}

For simplicity, we assume the transition between the two inflationary phases occurs instantaneously. A schematic illustration of the background and perturbed trajectories is shown in figure \ref{simplifiedtraj}.

In this case, for modes that exit the horizon before the transition, the total $\delta N$ evaluated at the end of inflation can be written as
\begin{align}
    &\mathcal{R}=\delta N=\delta N_1 +\delta N_2,
\end{align}
where $N_1\equiv n_*-n_k$ and $N_2\equiv n_{\mathrm{f}}-n_*$ are the total number of e-folds during the first and second stages of inflation, respectively. Here we introduce a new time variable $n(t)=\int_{t_i}^t Hdt$, $n_{*} $ and $n_\mathrm{f}$ represent the number of e-folds respect to an arbitrarily set initial time $t_i$ at the transition and at the end of inflation.

If we conduct the $\delta N$ calculation nonlinearly, one may observe a large deviation of the probability distribution function of $\mathcal{R}$ from Gaussianity, as observed already in single field inflation models \cite{Cai:2021zsp,Cai:2022erk,Pi:2022ysn,Kawaguchi:2023mgk,Wang:2024wxq}. 
Such non-Gaussian features are potentially detectable by upcoming observations. 
We leave this for future work. 

In this work, we restrict the calculation of $\delta N$ to linear level\footnote{See a more detailed analysis in \cite{Wang:2024vfv}}.  Assuming $\delta\chi_k$ and $\delta\phi$ are small perturbations, and treating both St-1 and St-2 as effective single-field inflation, we obtain the expression for $\delta N$:
\begin{align}
   \delta N= \left(\frac{\partial N_1}{\partial\phi}\delta\phi_k
% +\frac{\partial N_1}{\partial\chi}\delta\chi
\right)_{n_k}
+\left(\frac{\partial N_2}{\partial\chi}\delta\chi_k\right)_{n_{\star}},
\end{align}
where the first term is equivalent to the $\delta N$ in $R^2$ inflation, and the extra contribution of $\delta \chi$ appears in the second term. 
Intuitively speaking, the second term, namely the $\delta N_2$, reflects the perturbation of the number of e-folds during the second stage of inflation, induced by the mode $\delta\chi_k$ that exits the horizon during the first stage. 
This term is expected to be exponentially suppressed for large-scale modes since the amplitude of $\delta\chi_k$ evolves as 
\begin{equation}
    \delta\chi_k(n_\star)\approx\delta\chi_k(n_k)e^{-a(n_\star-n_k)/2},
\end{equation}
according to the Mukhanov-Sasaki equation. The decay rate $a$ is given by
\begin{equation}
    \begin{aligned}
    a&\equiv\mathrm{Re} \left(3-3\sqrt{1-\frac{16m^2_{\chi}}{9M^2}}\right)\\&=\mathrm{Re} \left(3-3\sqrt{1-\frac{16}{3}\xi}\right),
\end{aligned}
\end{equation}
during the St-1, which equals the maximum $3$ for the heavy $\chi$ case and smaller when decreasing $\xi$.

\begin{figure}[!htbp]
\centering
% \includegraphics[width=.48
% \textwidth]{transferfuncnew1.pdf}
\includegraphics[width=.6\textwidth]{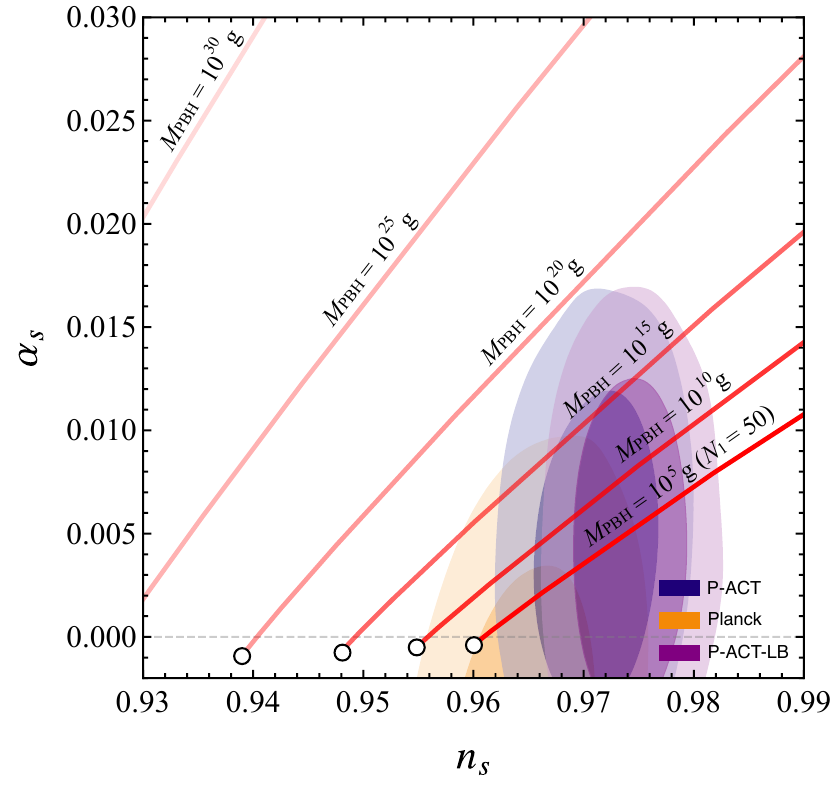}
\caption{\label{nsasana} The $n_s-\alpha_s$ relation shown in eq.~\eqref{nsas} for different $M_{\mathrm{PBH}}$s together with the joint constraints from Planck, P-ACT and P-ACT-LB (The constraint contours are taken from \cite{AtacamaCosmologyTelescope:2025nti}). To plot the figure, we use $\mathcal{P}_\mathcal{R}^{\mathrm{max}}/\mathcal{P}_\mathcal{R}^{\mathrm{CMB}}=10^{7}$ in order to have $f_{\mathrm{PBH}}=1$.
The hollow black dots refer to cases that $a=3$, or equivalently $\xi\geq 3/16$ when the small-scale amplification is well isolated from the large scales.}
\end{figure}
As a result, for modes that exit the horizon before the transition stage, the power spectrum contains both contributions from the dominating light field $\phi$ as well as the isocurvature field $\chi$. Assuming a constant $H$ and Bunch-Davis vacuum deep inside the horizon, the power spectrum for modes $k\leq k_1$, where $k_1$ is the mode that crosses the horizon during the end of the first phase of inflation (which is approximately the peak scale of the power spectrum), is given 
\begin{align}
    \mathcal{P}_{\mathcal{R}}(k)= \mathcal{P}_{\mathcal{R}}^{R^2}(k)+\mathcal{P}_{\mathcal{R}}^{\mathrm{peak}}\left(\frac{k}{k_1}\right)^{a}.
\end{align}
Here the $\mathcal{P}_{\mathcal{R}}^{R^2}$ refers to the single field $R^2$ model power spectrum, and the peak value of the spectrum $\mathcal{P}_{\mathcal{R}}^{\mathrm{peak}}$ can be approximate by a simple function of $\chi_0$, $\lambda$ and $m$. 

Accordingly, the spectral index, the running of the spectral index, and the tensor-to-scalar ratio at the CMB scale are approximately given by 
\begin{equation}
    \begin{aligned}
 &   n_s\approx1-\frac{2}{N_1}+\frac{\mathcal{P}_{\mathcal{R}}^{\mathrm{peak}}}{\mathcal{P}_{\mathcal{R}}^{\mathrm{CMB}}}e^{-aN_1}\left(\frac{2}{N_1}+a\right),\\
 &   \alpha_s\approx-\frac{2}{N_1^2}+\frac{\mathcal{P}_{\mathcal{R}}^{\mathrm{peak}}}{\mathcal{P}_{\mathcal{R}}^{\mathrm{CMB}}}e^{-aN_1}\left(\frac{2}{N_1^2}+\frac{2a}{N_1}+a^2\right),\\
  &  r\approx \frac{2}{N_1^2},
   \label{nsas}
\end{aligned}
\end{equation}
where $N_1$ is the number of e-folds before the end of the first stage of inflation corresponding to the CMB pivot scale. Since $k_1$ and $N_1$ correspond to the peak scale of the power spectrum, considering the sharp-peak case, they are related to the PBH mass by
% \begin{align}
%     M_{\mathrm{PBH}}\sim 3\times 10^{48}\mathrm{g} \times e^{-2N_1}.
% \end{align}
\begin{align}
    M_{\mathrm{PBH}}\sim 10^{18}\mathrm{g} \times e^{-2(N_1-35)}.
\end{align}
Therefore, to have PBH as all of the dark matter in the present universe, i.e., $f_{\mathrm{PBH}}=1$ is not yet constrained by observations, $N_1= 30\sim 40$ is needed \footnote{See reviews on the observational constraints on PBH as dark matter~\cite{Carr:2016drx, Sasaki:2018dmp, Carr:2020gox}}.

When the ratio of the power spectrum value at the peak scale and the CMB scale to be ${\mathcal{P}_{\mathcal{R}}^{\mathrm{peak}}}/{\mathcal{P}_{\mathcal{R}}^{\mathrm{CMB}}}\sim 10^{7}$ is fixed to generate a sizable abundance of primordial black holes today, i.e. $f_{\mathrm{PBH}}\sim 1$~\cite{Carr:2016drx,Green:2020jor,Sasaki:2018dmp}, if the coupling constant $\xi$ is larger than $3/16$ ($a=3$), the modifications from the blue-tilted component to the $n_s$ and $\alpha_s$ at CMB scale is exponentially suppressed, effectively decoupling small-scale features from large-scale observables.
In this limit, the large-scale power spectrum reduces to that of the single-field $R^2$ inflation, with $ \alpha_s\lesssim-0.0013$, $n_s\lesssim 0.95$ for $N_1\lesssim 40$.  When $e^{-aN_1}$ is of order $10^{-7}$ or less, the modification to the large-scale spectrum comes from the small-scale enhancement can be significant.
\begin{table}
    \centering
    \begin{tabular}{c|c|c|c}
    \hline%[2pt]
    Case & 1 & 2& 3\\
    \hline%[2pt] 
   \multicolumn{4}{c}{\textit{Parameters}}\\
    \hline%[2pt] 
    $M/M_{\rm pl}$& $1.35\times10^{-5}$ & $\ 1.78\times10^{-5}\ $ & $3.5\times10^{-5}$\\
      
    $m/M$ & $1/6$ & $\ 1/6\ $ & $1/6$\\
      
    $\xi$ & $0.0530$ & $0.0689$ & $0.1163$\\
     
    % $A\xi$ & $15/16$ & $1/3$ & $1/5$\\
     
    $\chi_0/M$ & $0.28$ & $0.39$ & $0.39$\\
    $\lambda$ & $4.7\times 10^{-12}$ & $2.9\times 10^{-12}$ & $6.8\times 10^{-12}$\\
   \hline%[2pt] 
   \multicolumn{4}{c}{\textit{CMB Observables}}\\
    \hline%[2pt] 
    $n_s(0.05\mathrm{Mpc}^{-1})$ & $0.975$ & $0.973$ & $0.973$\\
    $n_s(0.002\mathrm{Mpc}^{-1})$ & $0.966$ & $0.955$ & $0.912$\\
    $\alpha_s(0.05\mathrm{Mpc}^{-1})$ & $0.008$ & $0.014$ & $0.055,^{(a)}$\\
    \hline%[2pt] 
    \multicolumn{4}{c}{\textit{PBHs}}\\
    \hline%[2pt] 
    $M_{\mathrm{PBH}}/\mathrm{g}$ & $1.5\times10^{8}$ & $6\times 10^{18}$ & $1.7\times 10^{26}$\\
    $f_{\mathrm{PBH}}$ & $\sim1$ & $\sim1$ & $\ll 1$\\
    % $N_{\mathrm{CMB}}$ & $57$ & $53$ & $-1\times 10^{-4}$\\
    \hline%[2pt]
    \end{tabular}%
    \caption{The benchmark parameters and the observables. The primordial power spectrum is normalized by the Planck 2018 result  $\mathcal{P}_\mathcal{R}^{\mathrm{CMB}}=\mathcal{P}_\mathcal{R}(0.05\mathrm{Mpc}^{-1}) = 2.1 \times 10^{-9}$. A joint P-ACT-LB analysis reported a spectral index $n_{s}(0.05{\rm Mpc}^{-1})=0.9743 \pm 0.0034$. The total number of efolds $N$ of the inflation in our model can be tuned from $50\sim 60$ without affecting the IR by adjusting $\lambda$, thus not listed in the table. Notes: Excluded at more than $2\sigma$ level.}
    \label{tab:parameters}
\end{table}

\begin{figure*}[!htbp]
\centering
\includegraphics[width=1\textwidth]{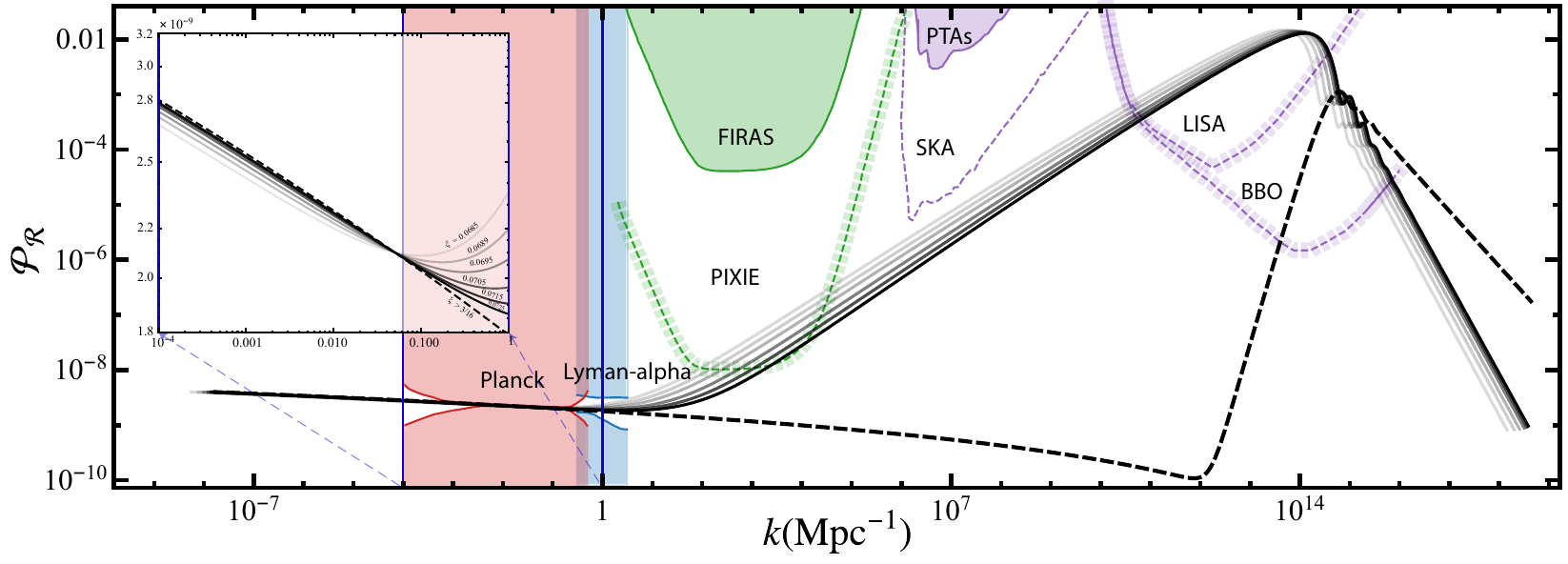}
\caption{\label{pscase2} The primordial power spectrum evaluated at the end of inflation shown alongside the (future) observational constraints. The black dashed line shows the result for $\xi>3/16$, whose power spectrum converges to the effective single-field $R^2$ inflation. The curves are obtained by numerically solving for the primordial power spectrum by the Transport method using the model parameters from Table~\ref{tab:parameters}, Case 2 except for a varying $\xi$. From the darkest to the lightest solid curves, we take  $ \xi=0.0725$, $0.0715$, $0.0705$, $0.0695$, $0.0689$ (Case 2), $0.0685$. The constraint plot comes from \cite{Kavanagh:2019}, including CMB constraints (red) from Planck~\cite{Planck:2018jri}, Lyman-$\alpha$ forest constraints~\cite{Zaldarriaga:2000mz} (blue),  $\mu$-distortion constraints (green) from FIRAS~\cite{Hu:1994bz}, and PIXIE (future)~\cite{Abitbol:2017vwa, Kogut:2011xw}, and gravitational wave constraints (purple) from PTAs~\cite{NANOGrav:2023gor,EPTA:2023fyk,Xu:2023wog}, SKA~\cite{Janssen:2014dka} (future), LISA~\cite{LISA:2017pwj} (future) and BBO~\cite{Harry:2006fi} (future).}
\end{figure*}
\section{Numerical results}
To account for the subtleties in the model, we perform numerical computations using the Transport method~\cite{Dias:2015rca}. 
% We listed 3 sets of benchmark parameters in Table. \ref{tab:parameters}, where set 1 for ultra-light PBHs, set 2 for dark matter PBHs and set 3 for weak lensing PBHs. All of the 3 sets of parameters can yield a slightly larger spectral index at the CMB scale $0.05/\mathrm{Mpc}$ which are consistent with the P-ACT result. 
We consider three sets of benchmark parameters, as listed in Table~\ref{tab:parameters}: Case 1 corresponds to ultra-light PBHs that evaporate before Big Bang Nucleosynthesis (BBN), Case 2 to asteroid mass PBHs within the dark matter window, and Case 3 to PBHs potentially detectable through microlensing surveys. All three parameter sets yield a slightly larger spectral index at the CMB pivot scale of $0.05\mathrm{Mpc}^{-1}$, which remains consistent with the P-ACT results.

However, Case 3 is excluded at more than the $2\sigma$ level based on the P-ACT-LB constraint on $\alpha_s$. 
This is due to the fact that, in our model, the production of PBHs with masses $\gtrsim 10^{20}\mathrm{g}$ is generally disfavored by ACT, as it is associated with a relatively large running of the spectral index (see figure \ref{nsasana}). As a result, we conclude that PBHs larger than $\sim 10^{20}$ g are excluded as a dark matter candidate in our model at 2 $\sigma$ level according to the P-ACT results.

Case 1 is not only consistent with current data but also phenomenologically intriguing. The ultra-light PBHs produced in this case are expected to evaporate before BBN via Hawking radiation~\cite{Hawking:1975vcx, Carr:1976zz, Hidalgo:2011fj}. Those evaporating tiny PBHs can help to reheat the universe, produce high-frequency gravitational waves by Hawking radiation, and induce second-order gravitational waves through the poltergeist mechanism~\cite{Dolgov:2000ht,Dolgov:2011cq,Dong:2015yjs,Inomata:2019ivs, Inomata:2020lmk,Papanikolaou:2020qtd,Domenech:2020ssp,Domenech:2021ztg,Ireland:2023avg,Domenech:2024wao,He:2024luf,Wang:2025lti}.
From quantum gravity point of view, those evaporated PBHs may abundantly leave remnants as dark matter~\cite{MacGibbon:1987my,Chen:2002tu,Lehmann:2019zgt,Domenech:2023mqk}.

In the following, we focus on the most compelling case, Case 2, where the near monochromatic asteroid mass PBHs produced from the model can constitute the whole dark matter~\cite{Carr:2020gox,Carr:2021bzv,Green:2024bam}, and is in excellent agreement with ACT observational constraints.

\begin{figure}[!htbp]
\centering
% \includegraphics[width=.48
% \textwidth]{transferfuncnew1.pdf}
\includegraphics[width=.6\textwidth]{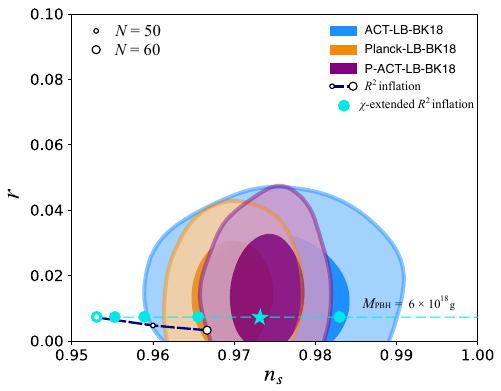}
\caption{\label{nsrcase2} The $n_s-r$ relation in the $R^2$ model and $\chi$-extended $R^2$ model shown alongside the joint constraints from BICEP/Keck observations~\cite{BICEP:2021xfz} (denoted as BK18) with Planck-LB (orange), ACT-LB (blue), and P-ACT-LB (purple). The constraint contours are taken from \cite{AtacamaCosmologyTelescope:2025nti}. The dark blue dashed line with hollow dots refers to the $n_s-r$ relation in the $R^2$ model, and the light cyan dashed line and markers represent the $\chi$-extended $R^2$ model with PBH production. For the light cyan plot, we use the model parameters from Table~\ref{tab:parameters}, Case 2 except for a varying $\xi$. From lower $n_s$ (left) to higher $n_s$ (right), the coupling constants used are $\xi=0.3125$ (hollow dot), $ 0.0725$, $0.0715$, $0.0705$, $0.0695$, $0.0689$ (star-shaped, Case 2), $0.0685$.}
\end{figure}

\begin{figure}[!htbp]
\centering
% \includegraphics[width=.48
% \textwidth]{transferfuncnew1.pdf}
\includegraphics[width=.6\textwidth]{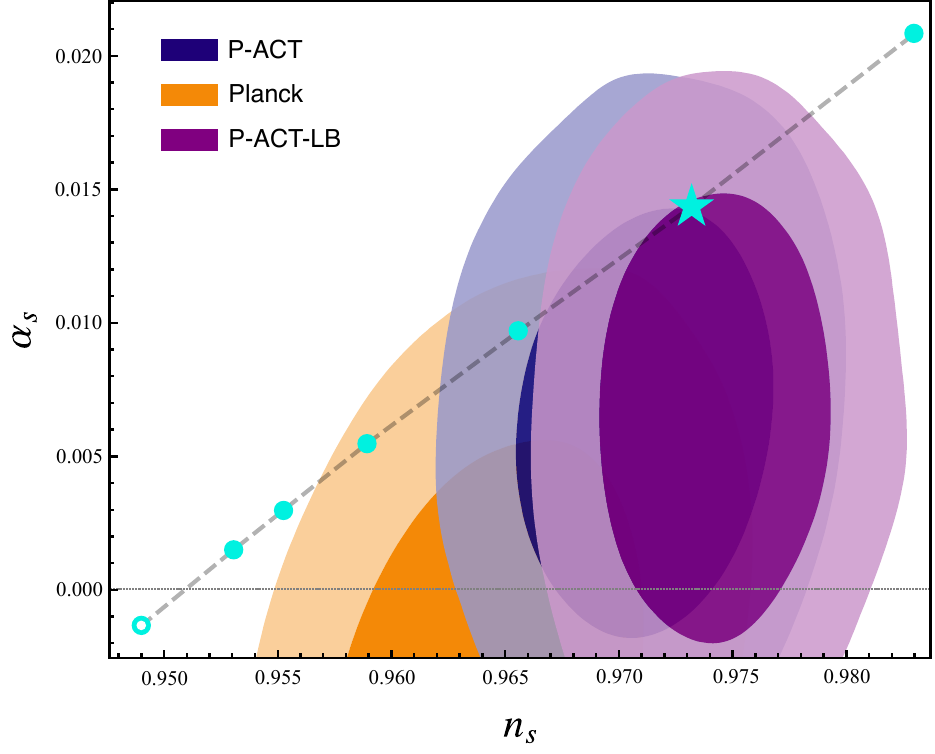}
\caption{\label{nsascase2} The $n_s-\alpha_s$ relation in $\chi$-extended $R^2$ model shown alongside the joint constraints from \textit{Planck}, P-ACT and P-ACT-LB (The constraint contours are taken from \cite{AtacamaCosmologyTelescope:2025nti}). The light cyan dots refer to the $\alpha_s-n_s$ relation in $\chi$-extended $R^2$ model with PBH production, for which we use the model parameters from Table~\ref{tab:parameters} \textbf{Case 2} except for a varying $\xi$.
From smaller $n_s$ (left) to larger $n_s$ (right), we take $\xi=0.3125$ (hollow dot), $ 0.0725$, $0.0715$, $0.0705$, $0.0695$, $0.0689$ (star-shaped, Case 2), $0.0685$.}
\end{figure}

The full power spectra resulting from parameter Case 2 with varying $\xi$s are shown in figure \ref{pscase2}. The variation of the spectral index and its running can be observed in figure \ref{pscase2}, and more intuitively in figure \ref{nsrcase2} and figure \ref{nsascase2}, where the tendencies are consistent with the analytical approximations given by eq.~\eqref{nsas}.

It is also worth mentioning that, as illustrated in figure \ref{pscase2}, the enhancement of the small-scale power spectrum not only affects the large-scale spectral index $n_s$ and its running $\alpha_s$, which can be probed through direct measurements of the primordial power spectrum from future CMB surveys~\cite{Abazajian:2019eic, SimonsObservatory:2018koc}, but also induces significant $\mu$-distortion~\cite{Chluba:2012gq,Chluba:2012we}—potentially detectable by PIXIE~\cite{Abitbol:2017vwa, Kogut:2011xw}\footnote{See detailed analysis in~\cite{Kim:2025dyi}.}.

Additionally, the associated induced gravitational waves fall within the sensitivity ranges of future detectors such as LISA~\cite{LISA:2017pwj}, DECIGO~\cite{Kawamura:2011zz}, Taiji~\cite{Hu:2017mde}/TianQin~\cite{TianQin:2015yph} and BBO~\cite{Harry:2006fi}. Thus, upcoming CMB and gravitational wave missions may provide critical observational tests of this scenario.

\section{Discussion and conclusion}

The recently released ACT data, combined with Planck observations, report a slightly larger scalar spectral index $n_s$ and a positive running $\alpha_s$ placing well-motivated $R^2$ inflation under tension at $2\sigma$ level. 
In this paper, we have shown that the tension can be resolved by introducing a non-minimally coupled scalar field $\chi$ to the $R^2$ model. 
By tuning the non-minimal coupling constant $\xi$, the model realizes a hybrid-like inflationary scenario, in which the non-minimally coupled scalar field $\chi$ contributes a blue-tilted component to the primordial power spectrum, yielding an increase in the $n_s$ and a small but positive $\alpha_s$, both consistent with current observational constraints.

At the same time, the blue-tilted power spectrum significantly enhances the fluctuation amplitude at small scales, potentially leading to the formation of primordial black holes (PBHs) that could constitute the entire dark matter abundance. In figure \ref{nsasana}, we illustrate the relationship between the spectral index $n_s$, its running $\alpha_s$, and the resulting PBH mass under the assumption that PBHs account for all dark matter.

Consistency with P-ACT observations implies that a positive running ($\alpha_s > 0$) may serve as indirect evidence for a blue-tilted spectrum and enhanced small-scale power, which can lead to the production of PBHs with masses $\lesssim 10^{20}~\mathrm{g}$ that constitute the entire dark matter content. Notably, the constraint from $\alpha_s-n_s$ allows partly the asteroid-mass window for PBH dark matter ($10^{17}\mathrm{g}\lesssim M_{\mathrm{PBH}}\lesssim10^{23}\mathrm{g}$) and the extended mass window if the memory burden effect is considered ($10^{5}\mathrm{g}\lesssim M_{\mathrm{PBH}}\lesssim 10^{10}\mathrm{g}$ when memory-burden parameter $n_{\mathrm{MB}}=2$~\cite{Alexandre:2024nuo,Thoss:2024hsr,Kohri:2024qpd}).

We emphasize that our model does not require PBHs to account for all dark matter in order to resolve the observed tension. Rather, it naturally correlates modifications to the large-scale spectrum with potentially observable small-scale features.

We also comment here on the potential observational signatures of the primordial blue-tilted power spectrum predicted by our model.

A slightly blue-tilted spectrum with a peak at small scales would be a distinctive prediction of this scenario. The blue tilt of the power spectrum can be probed by upcoming measurements of CMB spectral distortions, such as the $\mu$-distortion detectable by experiments like PIXIE. Meanwhile, the associated peak in the power spectrum can generate a second-order gravitational wave background, which may be within the reach of future detectors such as LISA~\cite{LISA:2017pwj}, DECIGO~\cite{Kawamura:2011zz}, Taiji~\cite{Hu:2017mde}/TianQin~\cite{TianQin:2015yph} and BBO~\cite{Harry:2006fi}. 

In addition, recent studies suggest that a blue-tilted primordial power spectrum at $k \gtrsim 1~\mathrm{Mpc}^{-1}$ could lead to the early formation of high-redshift dark matter halos~\cite{Hirano:2015wla, Parashari:2023cui,Hirano:2023auh}, and enhance the substructure abundance within low-redshift dark matter halos~\cite{Wu:2024tdh}. These features may help to explain the unexpectedly large number of massive galaxies observed at high redshift by the James Webb Space Telescope (JWST)~\cite{Gardner:2006ky,Labbe:2022ahb}, the 'too many satellite problem'~\cite{Kim:2017iwr,Muller2024_many_dwarf_satellites_M83} as well as the anomalous flux ratio problem~\cite{Mao:1997ek} detected in strong gravitational lensing systems.

Finally, we should stress the intriguing connection between the inflation scale and the small neutrino masses $m_{\nu_i}$. The $\chi$ boson has coupling to the right handed neutrinos as $g_i\chi N_{Ri}N_{Ri}$ which generates the large Majorana masses $M_i$ for the right-handed neutrinos after the total inflation ends as explained in the introduction. Through the seesaw mechanism~\cite{Minkowski:1977sc,Yanagida:1979as,Yanagida:1979gs,Gell-Mann:1979vob}, the observed neutrino masses $m_{\nu_i}$ are given by $m_{\nu_i}\simeq h_i^2\langle H\rangle^2/M_i$. Here, the $H$ is the standard-model Higgs boson, $M_{i} \simeq g_i\langle\chi\rangle$,  and the $\langle\chi\rangle=v$ is determined by the inflation scale as shown in the text. Therefore, it is very interesting that there is a strong correlation between the inflation scale and the observed small neutrino masses. 

\begin{acknowledgments}
We thank Misao Sasaki and Ying-li Zhang for useful discussions. 
This work is supported in part by JSPS KAKENHI Grant Nos. JP20H05853, JP24K00624 [X.W.]; JP23KF0289, JP24H01825, JP24K07027 [K.K.]; No. JP24H02244 [T.T.Y.], and by Forefront Physics and Mathematics Program to Drive Transformation (FoPM), a World-leading Innovative Graduate Study (WINGS) Program, the University of Tokyo [X.W.].
Kavli IPMU is supported by World Premier International Research Center Initiative (WPI), MEXT, Japan.
X.W. would also like to acknowledge the insightful input from the 'Random Discussion' and the 'Workshop on the Physics and Mathematics of the Universe' held at Kavli IPMU, which greatly contributed to the completion of this work.
% K.K. is supported by KAKENHI Grant No. JP23KF0289, No. JP24H01825, and No. JP24K07027.  
\end{acknowledgments}

% \appendix
% \section{Effect of the $R^3$ term on large-scale power spectrum }
% We consider the extended $R^2$ model with a cubic curvature term,
% \begin{align}
%      S_{J}=\int d^4x\sqrt{-g}\ \frac{1}{2}M_{\mathrm{pl}}^2f(R,\chi),
% \end{align}
% where
% \begin{align}
%     f(R)= R 
%     + \frac{R^2}{6M^2}
%     + q \frac{R^3}{3M^4}
% \end{align}
% , here $q$ is a dimensionless parameter.

\appendix
\bibliography{apssamp}% Produces the bibliography via BibTeX.
\bibliographystyle{apsrev4-1}
\end{document}